\documentclass[journal]{IEEEtran}

\usepackage[utf8]{inputenc}
\usepackage{color}
\usepackage{xcolor}
\usepackage{array}
\usepackage{verbatim}
\usepackage{float}
\usepackage{amsmath}
\usepackage{amsthm}
\usepackage{amssymb}
\usepackage{graphicx}
\usepackage{longtable}
\usepackage{multirow}
\usepackage{booktabs}
\usepackage[unicode=true,
bookmarks=false,
breaklinks=false,pdfborder={0 0 1},colorlinks=false]
{hyperref}
\hypersetup{
	colorlinks,bookmarksopen,bookmarksnumbered,citecolor=blue,urlcolor=blue}
\usepackage{cite}
\usepackage{array}
\usepackage{overpic}
\usepackage{dblfloatfix}
\usepackage{balance}
\usepackage{lipsum}
\usepackage{mathtools}
\usepackage{cuted}

\usepackage{algorithmic}
\usepackage{longtable}

\floatstyle{ruled}
\newfloat{algorithm}{tbp}{loa}
\providecommand{\algorithmname}{Algorithm}
\floatname{algorithm}{\protect\algorithmname}

\makeatletter
\let\oldforeign@language\foreign@language
\DeclareRobustCommand{\foreign@language}[1]{%
	\lowercase{\oldforeign@language{#1}}}

\let\oldforeign@language\foreign@language
\DeclareRobustCommand{\foreign@language}[1]{%
	\lowercase{\oldforeign@language{#1}}}

\ifCLASSINFOpdf
\else
\fi

\@ifundefined{showcaptionsetup}{}{%
	\PassOptionsToPackage{caption=false}{subfig}}
\usepackage{subfig}

\usepackage{balance}

\ifCLASSINFOpdf
\else
\fi

\ifCLASSINFOpdf
\else
\fi

\def\ps@IEEEtitlepagestyle{%
	\def\@oddhead{\parbox[t][\height][t]{\textwidth}{\centering \scriptsize
			Personal use of this material is permitted. Permission from the author(s) and/or copyright holder(s), must be obtained for all other uses. Please contact us and provide details if you believe this document breaches copyrights.\\
			\noindent\makebox[\linewidth]{}
		}\hfil\hbox{}}%
	\def\@evenhead{\scriptsize\thepage \hfil \leftmark\mbox{}}%
	\def\@oddfoot{\parbox[t][\height][l]{\textwidth}{
			\vspace{-20pt}{\rule{\textwidth}{0.4pt}}\\ \footnotesize{\bf{\footnotesize\textcolor{red}{Hasan, H. "A cost effective deaf-mute electronic assistant system using myo armband and smartphone." International Journal of Science and Research (IJSR) 6 (2017): 950-954.}}}\\
			\noindent\makebox[\linewidth]
		}\hfil\hbox{}}%
	\def\@evenfoot{\MYfooter}}

\makeatother
\begin{document}
	\bstctlcite{IEEEexample:BSTcontrol}
							
\title{A Cost Effective Deaf-mute Electronic Assistant System Using Myo Armband and Smartphone}

\author{Hussein Naeem Hasan
	\thanks{H. N. Hasan is with the Department of Mechatronics, Al-Khwarizmi College of Engineering, University of Baghdad, Karrada Kharij Road, Al-Jadyria, Baghdad, Iraq, (hnaeem@kecbu.uobaghdad.edu.iq).}
}

\maketitle
\begin{abstract}
Communication is essential feature in human communities. For some reasons, deaf-mute disabled people lose their ability to hear, speak, or both which makes them suffer to communicate and convey their ideas, especially with normal people. Sign language is the solution for communication in the deaf-mute societies, but it is difficult for the rest of people to understand. Therefore, in this work, a cost effective Deaf-Mute Electronic Assistant System (DMEAS) has been developed to help in solving the communication problem between the deaf-mute people and the normal people. The system hardware consists only of a Myo armband from \textit{Thalmic Labs} and a smartphone. The Myo armband reads the electromyographic signals of the muscles of the disabled person's forearm through non-invasive, surface mounted electrodes (sEMG) and sends the data directly to the smartphone via Bluetooth. The smartphone will recognize and interpret them to a predefined word depending on the hand gestures. The recognized gesture will be displayed on the smartphone screen as well as displaying the text of the word related to the gesture and its voice record. All the EMG signals are processed using discrete wavelet transform and classified by neural network classifier. The system was extensively tested through experiments by normal subjects to prove its functionality. 
\end{abstract}

\section{Introduction}\label{intro}
\IEEEPARstart{R}{ecently} , technology has entered the "smart" era. We can see smart devices or systems everywhere in our daily lives, for instance, smartphone, smart TV, smart car, and so on. This dramatic advance in technology can be exploited to serve the humanity. We can use it to help people who have disability to regain "even a part of" their ability to get involved in the daily lives and join their society. Deaf-mute people have disability of hearing loss, speaking loss, or both. Researchers have done many valuable works to solve the communication problem between the deaf-mute people and normal people \cite{ ali2016hand, Pallavi, Sunita, padmanabhan, Shreyashi, Rohit, Lillian}. Depending on the sign language, researchers have employed the gesture recognition techniques to interpret the hand and fingers' gestures to letters or some sort of words. In some works, glove-based techniques are used to obtain the gestures data in order to convert them to readable letters \cite{ali2016hand, Pallavi, Sunita, padmanabhan}. The glove contains five flex sensors to measure the finger movements as well a motion sensor to measure the hand orientation. The flex sensor changes its resistance depending on the bending amount on its length which is calibrated to the movement of the corresponding finger. Although glove-based techniques are reliable and require non-complicated processing burden, they are uncomfortable to wear and cause natural motion difficulties for the user. 
Seeking other methods, vision-based techniques have been used to capture the hand motion and gestures. In these techniques, cameras are used to capture a live video for the hand and send it to a processing unit in order to extract and recognize the gestures and distinguish between them using image processing\cite{Shreyashi, Rohit, sohalreview}. Vision-based technique suffers some serious drawbacks due to the fact that it is susceptible to the environmental effects such as light intensity and background conditions \cite{ali2016hand}. In addition, it requires the user to be in a specific position relative to the camera which makes the user lacks the freedom to move around. 
On the other hand, electromyography (EMG)-based approach is used to read the hand gestures through surface mounted electrodes (sEMG) sensors. The later measure the electrical activities of the muscles and extract the electromyographic signals \cite{Ahsan, Chen11, Kim}. These signals can be processed and classified for gestures recognition. Thalmic Labs employed the sEMG sensors to design the Myo armband for an accurate and reliable gesture control as a wearable technology \cite{Thalmic}. Al-Tunawi \cite{Lillian} used the Myo armband to design a system to help interpreting the sign language gestures; however the system requires the Myo armband to be connected to a personal computer to receive and process the data. Therefore, the system has some limitation like, delay in real-time proceeding due to the multiple sub-processing units as well as high cost and difficulty to the user to use it outside due to the personal computer requirement.  In our work, we used the low cost Myo armband for the EMG-based gestures recognition and a smartphone only for the same purpose which is a cost and size effective in comparison with other systems. The Myo armband accurately reads the gestures and sends the data to the smartphone in order to be recognized, displayed in text, and audibly spoken up.

The paper is organized as follows: Section \ref{system_components} provides detailed descriptions of the system components 
, including the Myo armband, smartphone, and the application. Section \ref{gestures} presents the gestures definition and thier related words and phrases. The signal processing and classification are presented in Section \ref{classification}. Finally, the conclusion is stated in Section \ref{Conclusion}.

\section{DMEAS System Components \label{system_components}}
Cost, size, and simplicity are very effective factors in any project. Therefore, we kept these factors in mind as the first step in designing our system. The system simply consists of two main hardware components: a Myo armband and a smartphone as shown in Figure \ref{Assembly}. As we mentioned in the title and introduction that our system is cost effective, table \ref{table:t1} shows the system components and their costs. In this section, we will give a detailed description for the main components in the system and explain their main functions:
\begin{figure}[btph!]
    \centering
        \includegraphics[width=0.45\textwidth]{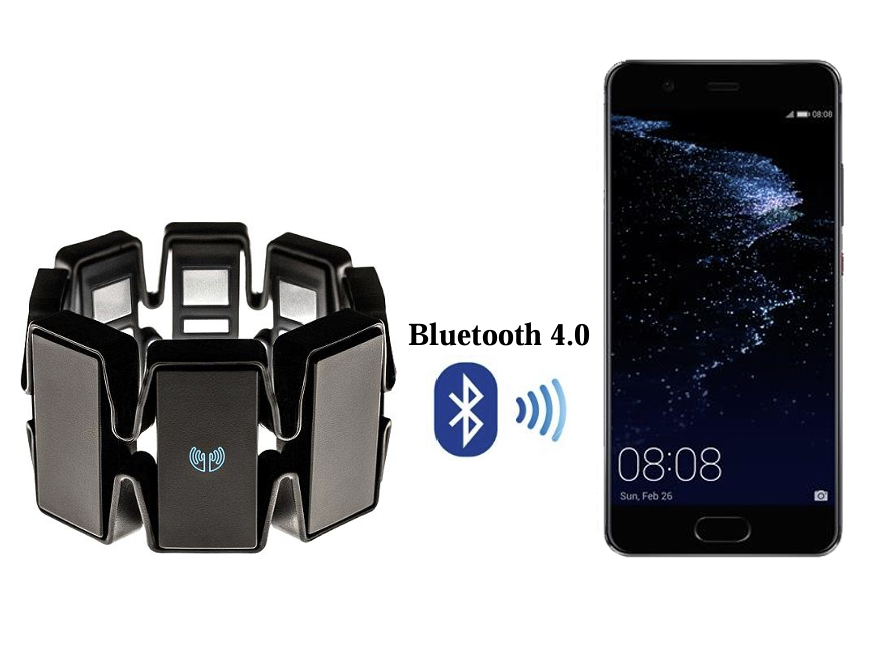}
    \caption{The Main System Components.}
    \label{Assembly}
\end{figure}

\begin{table}[h]
\caption{Components cost in US dollar.}
\centering
\begin{tabular}{lll}
\hline
No. & Component & Price (\$) \\
\hline
1 & Myo Armband & 200 \\
2 & Smartphone & less than 100 \\
3 & Application & Free \\
\hline
\end{tabular}
\label{table:t1}
\end{table}

\subsection{The Myo Armband \label{myo_armband}}
\begin{figure}[h!]
    \centering
        \includegraphics[width=0.56\textwidth]{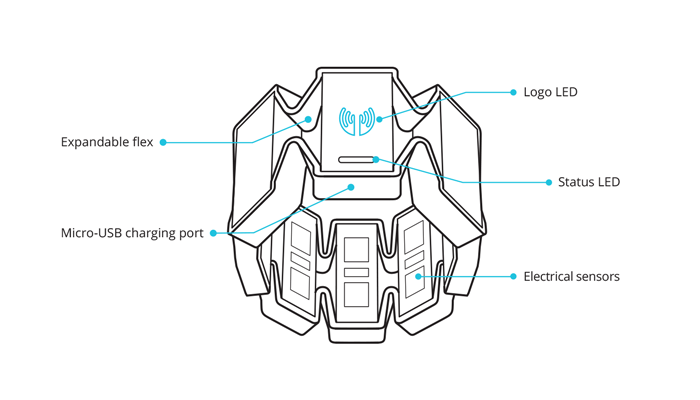}
    \caption{Myo Armband Gesture Control \cite{Thalmic}.}
    \label{Myo}
\end{figure}

Figure \ref{Myo} shows the Myo armband and its main components which is designed by Thalmic Labs as a wearable device. It contains eight non-invasive, surface mounted bio-sensors to read the electromyographic signals of human forearm's muscles. In addition, it contains a nine axes Inertial Measurement Unit (IMU) that consists of a three axis accelerometer, three axis gyroscope, and three axis magnetometer. The IMU unit is responsible for measuring the motion and movement of the armband and gives it in the form of orientation and acceleration vector \cite{Thalmic}.
Moreover, it contains a Bluetooth 4.0 Low Energy (BLE) for communication and a haptic sensor. All of these components are controlled by ARM Cortex M4 processor and powered by a small rechargeable lithium battery that is recharged through a mini USB port. The entire system is encapsulated inside eight plastic casing. The band contains stretchable parts that connect the eight segments together and give the band flexibility to fit the user's arm. The logo LED has two states pulsing and solid to show the synchronization of the armband and the status LED shows the connection and charging status. The Myo armband can be used easily with Windows and Mac operation systems for computers and Android and iOS for smartphones and tablets \cite{Thalmic}.

\begin{figure*}
    \centering
        \includegraphics[width=0.9\textwidth]{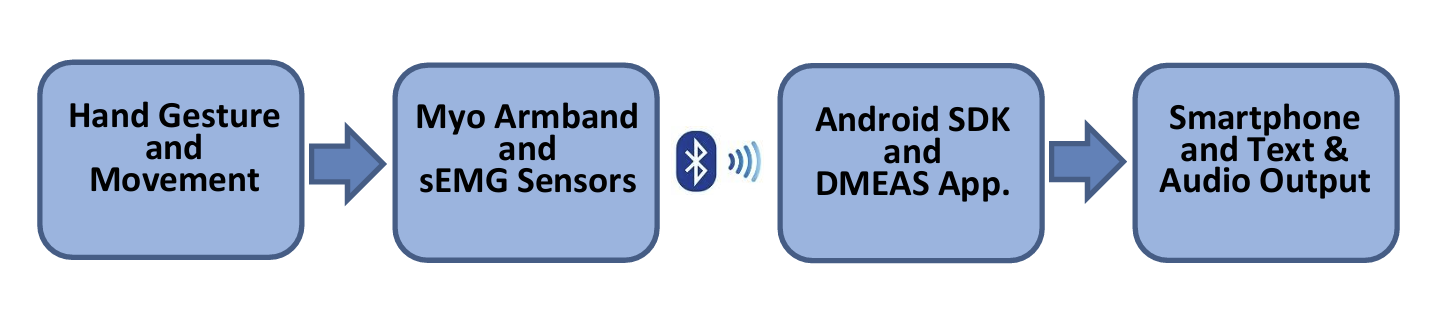}
    \caption{The Block Diagram of the DMEAS System.}
    \label{scheme}
\end{figure*}

The main function of the Myo armband in this project is to measure the electrical activities of the user muscles through its eight bio-sensors and process the data as well as send them to the smartphone via Bluetooth as shown in Figure \ref{scheme}.

\subsection{The Smartphone}
\label{components}
As mentioned above, the Myo armband is compatible with Android, so an Android smartphone was used in this work. We chose to use the smartphone in our system for the following reasons:

\begin{enumerate}
\item It is a compatible, easy to use smart system.
\item It contains many useful components on one small-sized platform that can be easily accessed and used in our project such as high powered processing unit, touch display screen, audio system, memory, and many other features that make it an excellent choice for many smart projects.
\item It is widely used and portable and can be found everywhere.
\item It has low cost; many Android devices are relatively cheap.
\end{enumerate}

For the project's purposes, an Android application has been developed to communicate with the Myo armband via Bluetooth. The application receives the data from the Myo armband and does the following tasks depending on its database:
\begin{enumerate}
\item  Distinguishes between the received gestures' signals and identifies the specific one to take the related actions.
\item  Displays a graph of the gesture on the screen as well as the text of a specified word related to the particular gesture.
\item  Operates a predefined human voice record of the specified word.
\end{enumerate}

The application has been developed using the Android Studio IDE environment from \textit{Google} depending on the Software Development Kit (SDK) for the Android form the \textit{Thalmic Labs} \cite{Thalmic} as shown in Figure \ref{scheme}.

\section{Gestures Definitions \label{gestures}}
\begin{figure}[h!]
    \centering
        \includegraphics[width=0.42\textwidth]{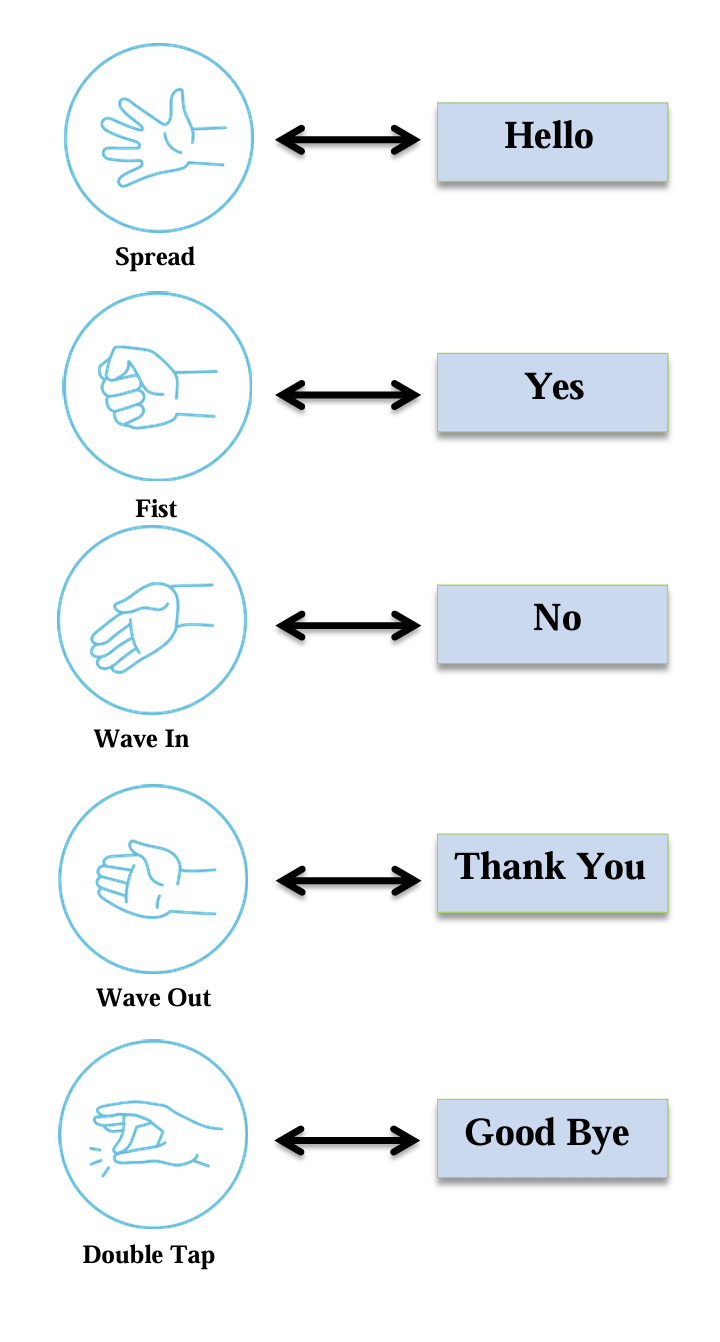}
    \caption{Gestures and their relative words.}
    \label{Gestures}
\end{figure}

In the literature, we have noticed that many researchers focused on interpreting the sign language gesture related to the letters \cite{ali2016hand, Pallavi, Sunita, padmanabhan, Shreyashi, Rohit, Lillian}. Letters themselves have no specific meaning compared to entire words. Following this approach makes it relatively complex to speak a complete word, especially the long words which consist of many letters and require many gestures and high computational power. In opposite, in our project, we used a system of relating every gesture with a predefined specific word or phrase to give a clear idea as shown in Figure \ref{Gestures}. The gestures and their related words or phrases here are just for example to prove the concept and can be modified in accordance to the user's needs and language.

\section{Signal Processing and Classification \label{classification}}
\begin{figure}[tb!]
    \centering
        \includegraphics[width=0.51\textwidth]{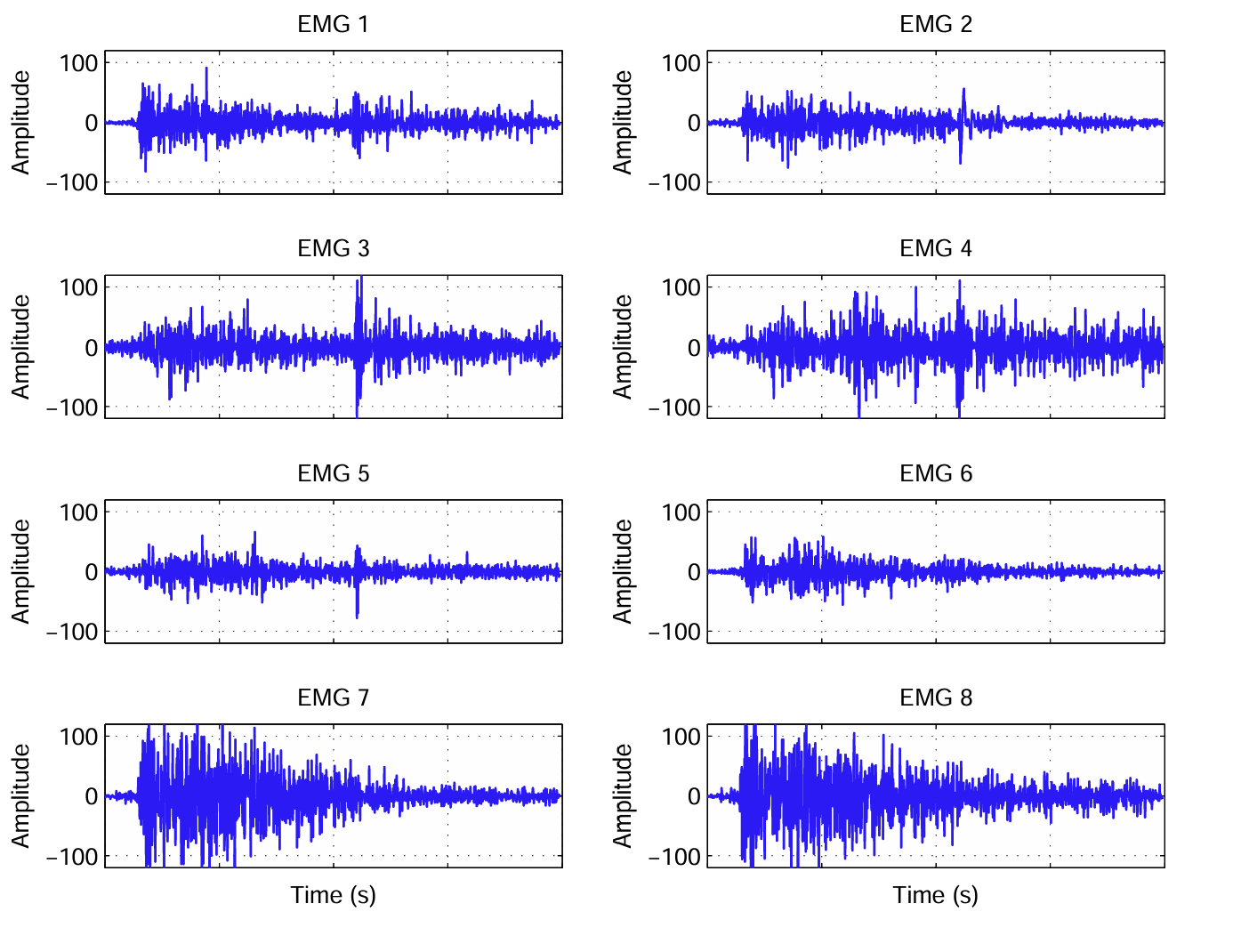}
   \caption{A sample of the EMG signals.}
    \label{diag}
\end{figure}

As mentioned in section \ref{myo_armband}, the Myo armband contains eight bio-sensors to measure the electric activity signals of the user arm's muscles (EMG signals). Figure \ref{diag} shows a sample of the EMG signals from the eight bio-sensors of the Myo armband.  The raw EMG signals from the Myo armband are processed to remove the low and high frequencies noises and the powerline noise (50/ 60 Hz). After the original EMG signal were retrieved, features extraction operations were applied to extract the distinctive characteristic of the EMG signals.  Due to the non-stationary nature of the EMG signal, discrete wavelet transform was used for the feature extraction. Using the discrete wavelet transform produces a very large feature coefficients vector which will require high computational burden. Therefore,  we used features reduction statistical technique in order to reduce the dimension of the feature vector. In this technique, four statistic coefficients of means ($\bar{\eta}$), variance ($\sigma$), skewness ($\gamma$), and kurtosis ($\kappa$) are calculated which will provide huge reduction in the dimension of the feature vector and the computational power and time. The means, variance, skewness, and kurtosis are given by Eq. \eqref{equ1} to \eqref{equ4} respectively. 
\begin{align}
    \bar{\eta} & =\frac{1}{N}\sum\limits_{n=1}^N (s[n]) \label{equ1} \\
    \sigma & =\Bigg(\frac{1}{N}\sum\limits_{n=1}^N (s[n]-\bar{\eta})^2\Bigg)^{1/2} \\
    \gamma & =\frac{1}{N\sigma^3}\sum\limits_{n=1}^N (s[n]-\bar{\eta})^3 \\
    \kappa & =\frac{1}{N\sigma^4}\sum\limits_{n=1}^N (s[n]-\bar{\eta})^4 \label{equ4}
\end{align}

Where the $s[n]$ is the EMG signal and $N$ is the number of samples. The features vector that extracted in the previous step are used as an input vector to the Neural Network (NN) classifier. We used feedforward multi-layers perceptrons (MLP) with log-sigmoid transfer functions to build our classifier that was trained using backpropagation algorithm.  All the signal processing, training and classification were done using MATLAB (2014) then implemented in the Android Application.    

\section{Results}
\begin{figure}[b!]
    \centering
        \includegraphics[scale=0.35]{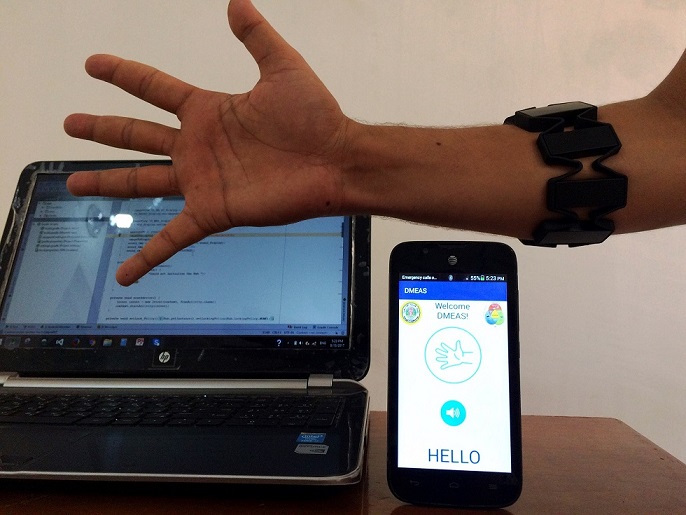}
    \caption{The Gesture "Fingers Spread" with its Related Word "HELLO" on the Smartphone Screen.}
    \label{Result1}
\end{figure}

As mention in Section \ref{components}, the Android application has been developed to manipulate the data that received from the Myo armband. Depending on predefined database and the classification technique that explained in Section \ref{classification}, the application identifies the received hand gesture then displays its graph, the text of its related word or phrase , and an audio icon on the smartphone screen for one second of time as well as plays a voice record for the related text. After that, the application removes all the displayed items from the screen in order to be ready to the next process.
The system was extensively tested by a normal person, and then it was easily used to train disabled people. The results were as follows: as shown in Figure \ref{Result1}, we can see that a trained user wears the Myo armband with an android smartphone. In this figure, the user made the "Fingers Spread" gesture with his hand, and the smartphone displayed the graph of the gesture with its related word "HELLO".

\begin{figure}[ht]
    \centering
        \includegraphics[scale=0.35]{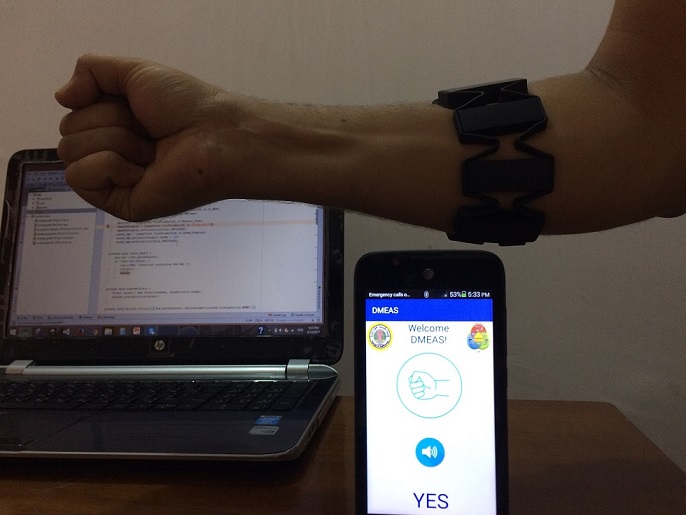}
    \caption{The Gesture "Fist" with its Related Word "YES" on the Smartphone Screen.}
    \label{Result2}
\end{figure}

\begin{figure}[h!]
    \centering
        \includegraphics[scale=0.37]{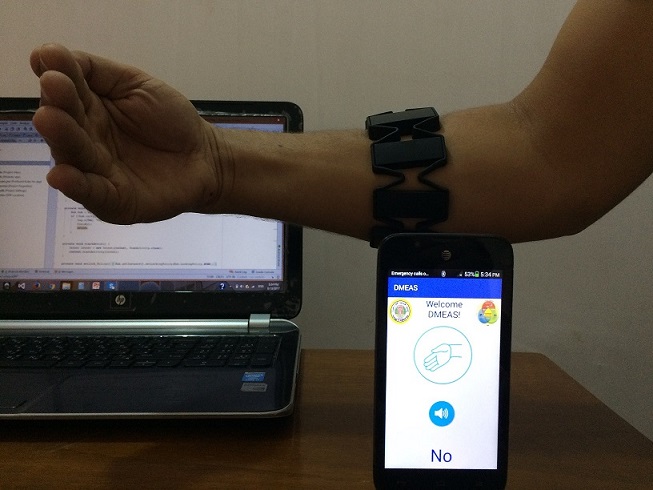}
    \caption{ The Gesture "WaveIn" with its Related Word "NO" on the Smartphone Screen.}
    \label{Result3}
\end{figure}

\begin{figure}[h!]
    \centering
        \includegraphics[scale=0.355]{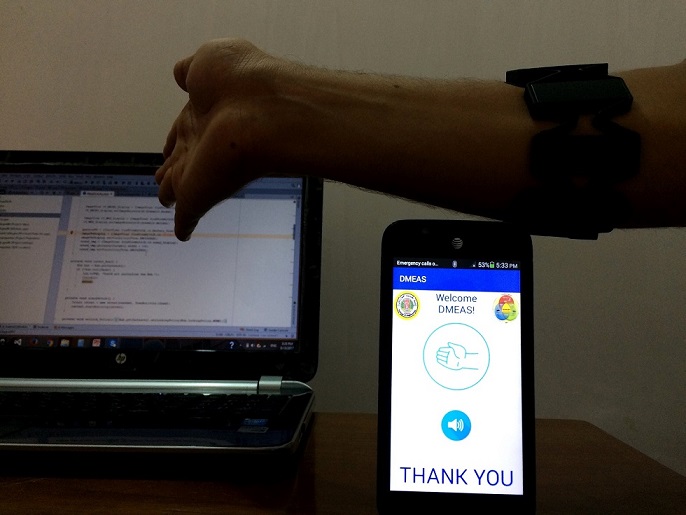}
    \caption{The Gesture "WaveOut" with its Related phrase "THANK YOU" on the Smartphone Screen.}
    \label{Result4}
\end{figure}

\begin{figure}[h!]
    \centering
        \includegraphics[scale=0.373]{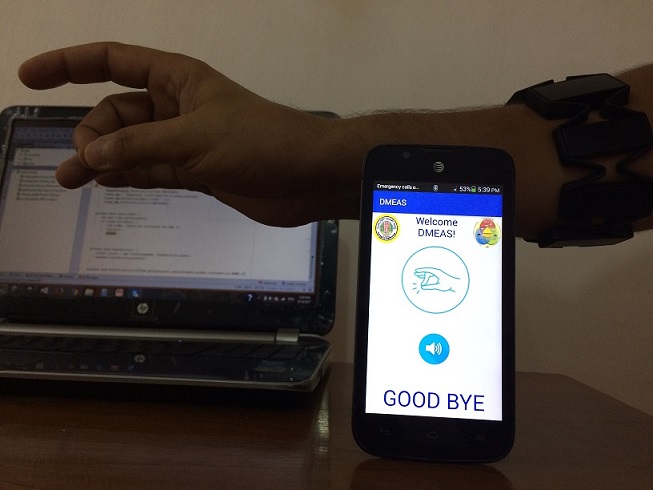}
    \caption{The Gesture "DoubleTap" with its Related phrase "GOOD BYE" on the Smartphone Screen.}
    \label{Result5}
\end{figure}

Figures \ref{Result2} to \ref{Result5} show other gestures , "Fist", "WaveIn", "WaveOut", and "DoubleTap"  with their related words or phrases "YES", "NO", "THANK YOU", and "GOOD BYE" respectively, that were made by the user on the smartphone screen. Depending on its confogration, the neural network classifier was able to perfectly recognize the received gesture patterns.

\section{Conclusion\label{Conclusion}}
Conveying ideas and special need for the deaf-mute people is difficult to be understood without interpreter help. Using technology to help them to communicate with normal people makes their life relatively easy to express their needs. In this work, the DMEAS has been developed and tested to convert the hand gestures of a disabled user to audible and text outputs. The system has small size and low cost with the use of the user's smartphone which can be within reach wherever the user goes. With its free application, it only requires a simple training to update the database in order to start talking with normal people.

\balance
\bibliographystyle{IEEEtran}
\bibliography{reference}

\end{document}